\documentclass[smallextended]{svjour3}       
\smartqed  
\usepackage{graphicx}
 \journalname{Int J Theor Phys}
\begin{document}

\title{Chromomagnetism in nuclear matter} 
\author{Akhilesh Ranjan \and P. K. Raina}
\institute{Akhilesh Ranjan \at
Department of Physics and Meteorology, Indian Institute of Technology Kharagpur, WB, India, 721302
\email{akranjanji@rediffmail.com}   
           \and
                P. K. Raina \at
Department of Physics and Meteorology, Indian Institute of Technology Kharagpur, WB, India, 721302
\email{pkraina@phy.iitkgp.ernet.in}   
             \emph{Present address:} 
Department of Physics, Indian Institute of Technology Ropar, Rupnagar, Punjab, India, 140001 
}

\date{Received: date / Accepted: date}

\maketitle

\begin{abstract}
 Quarks are color charged particles. Due to their motion there is a strong 
possibility of generation of color magnetic field. It is shown that however 
hadrons are color singlet particles they may have non-zero color magnetic 
moment. Due to this color magnetic moment hadrons can show color interaction. 
In this paper we have studied the chromomagnetic properties of nuclear matter.

\end{abstract}

\section{Introduction}
\label{intro}
Quark-gluon are color charged particles. The dominant interaction among quarks 
and gluons are strong(color) interaction. Color interactions are analogous to 
electromagnetic interactions. Color charge particles produce chromoelectric 
field. Therefore chromomagnetic field will also be produced due to the motion 
of colored particles. Effect of chromomagnetism was earlier studied by Rujula, 
Georgi, and Glashow to explain mass splitting of hadron spectrum \cite{georgi}. 
In this work they also predicted the existence of new states of hadrons with 
charm quarks and the mass of charm quarks. Chromomagnetism was also found 
useful to explain decay as well as production mechanisms of many hadrons 
\cite{bourdeau,aramayo}. In these works various kind of chromomagnetic forces 
were found and were supported by experimental results. 

In a separate work Sapirstein examined the quark color magnetic moment in the 
framework of a quark with a constant external color magnetic field 
\cite{sapirstein}. Ninomiya and Sakai studied the chromomagnetism to 
understand the phase transitions in quantum chromodynamics(QCD) 
\cite{ninomiya}. Kapusta has studied the phase transition in a system of 
gluons interacting with a constant color magnetic field \cite{kapusta}. 
Therefore we find that chromomagnetism has played a vital role in explaining 
some complicated problems of QCD.     

Color charge on quarks and gluons are responsible for strong or 
non-abelian interactions. Because of color charges they may produce 
chromoelectric and chromomagnetic fields. The chromoelectrodynamics 
of these particles is governed by the Yang-Mills equations which are 
analogs of the Maxwell's equations. As we know that hadrons are color 
neutral particles but with the help of string model we have proved that 
hadrons can have color magnetic moment. Because of color magnetic moment 
hadrons may show color interactions. In section two we have calculated 
the color magnetic moments of mesons and baryons separately. Then we have 
calculated the partition function. With the help of partition function 
we have studied the chromomagnetic properties of nuclear matter and 
possibilities of phase transition.

\section{Formalism}
Quarks and gluons are color charged particles. Due to color charge 
and their motion they can produce chromoelectric and chromomagnetic 
fields. In this paper we have discussed only quarks' contribution in 
chromomagnetism. Gluonic contributions will be discussed later in some 
other work.  

Hadrons are color neutral particles but they can have color magnetic moment 
like a neutron, which is electrically neutral particle but it has non-zero  
magnetic moment. Therefore hadrons can have color interaction due to their 
color magnetic moment. Hence we have first estimated the chromomagnetic 
moment of hadrons. We have done the formulation in the framework of the 
string model of hadrons. According to this quark(or antiquark) are attached 
at the ends of the string. The string rotates about its center of mass. Due 
to rotation color currents are generated. These color currents are of two 
type: (i) abelian, and (ii) non-abelian. The abelian current is generated 
due to the orbital motion of quarks and non-abelian current is generated due 
to the temporal variation of color charge which is well described by the Wong 
equation \cite{wong}. The Wong's equation is written as 
\begin{eqnarray}
\label{wo} 
\frac{dQ^a}{d\tau}=f^{abc}u_{\mu}Q^bA^{c\mu}
\end{eqnarray}
where $A^{a\mu}$ is the gauge potential, and $f^{abc}$ is the structure 
constant of the gauge group, $Q^a$ is the color charge of a quark, and $a$ 
is the color index. $u_{\mu}$ are four velocity vectors given by 
$u_{\mu}=\gamma(1,-\vec{v})$. It should be noted that in eq\ref{wo} the 
derivative is {\it wrt} $\tau$(proper time) not {\it wrt} `$t$'. 
Therefore the color current produced by a quark/antiquark is given by  
\begin{eqnarray} 
I^a_{total}=I^a_{abelian}+I^a_{non-abelian} 
\end{eqnarray} 
With the help of Wong equation the color current is obtained as 
\begin{eqnarray} 
I^a_{total}=\frac{Q^a\omega}{2\pi}+\frac{f^{abc}}{\gamma}Q^bu_{\mu}A^{c\mu} 
\end{eqnarray} 
where $\omega$ is the angular velocity of quarks and $\gamma$ is the 
relativistic factor. Due to these currents hadrons have intrinsic color 
magnetic moments. 

 \begin{figure}[htb]
\vspace*{5mm}
\hspace*{0mm}
\includegraphics[height=6.0 cm, width=10.0 cm]{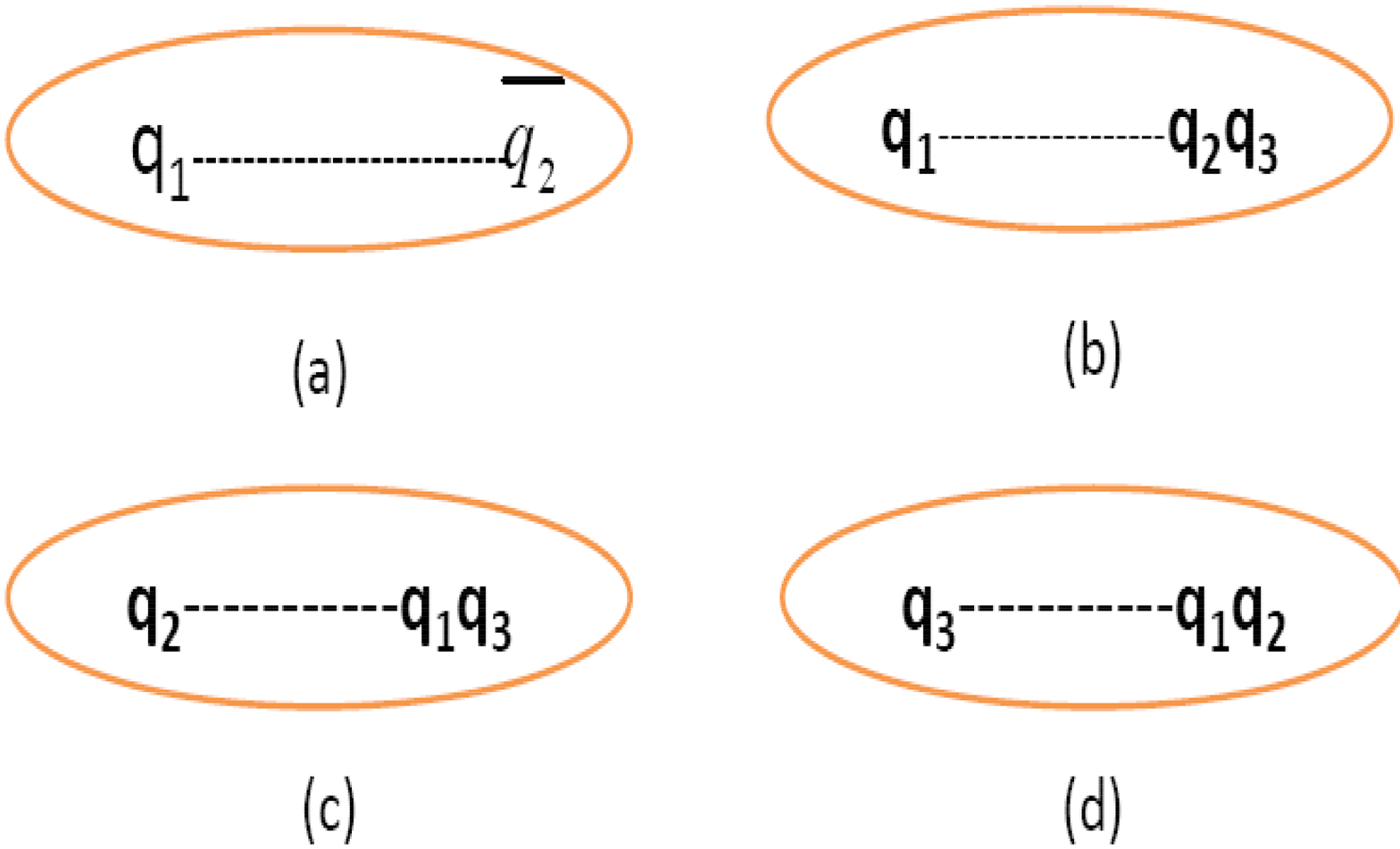}
\vspace*{-0mm}
\caption{(a) string model of meson; Fig~(b),(c),(d): string model of baryon 
with different but equally probable configurations.}
\label{hadron}
\end{figure} 

We have divided our analysis into two parts: (i) mesons $\&$ (ii) baryons. 
\subsection{Color magnetic moment of mesons} 
A meson contains quark and antiquark pair at the end of flux string 
(see fig a). Let us consider a general case of meson which contains two 
different kinds of quark-antiquark. Due to color neutrality of meson 
quark-antiquark pair will have equal and opposite color charges. The orbital 
color magnetic moment($\mu_1^a$) associated with quark $'1'$ is given by 
\begin{eqnarray} 
\label{q1-mag-mom} 
\vec{\mu_1}^a=\left[\frac{Q^a\omega}{2\pi}+f^{abc}Q^b\{A_o^c(\vec{r}_1)-\vec{v}_1\cdot\vec{A}^c(\vec{r}_1)\}\right]\frac{\pi l^2m_2^2}{(m_1+m_2)^2}\hat{n}
\end{eqnarray} 
where $Q^a$, $m_1$ are color charge and mass of quark $'1'$, 
$A_o^c \& \vec{A}^c$ are gauge fields at $\vec{r}_1$. It is assumed that 
the quark $q_1$ is situated at location $\vec{r}_1$ ${\it wrt}$ center of 
mass of meson. $f^{abc}$ is the structure constant of SU(3) group. $m_2$ is 
the mass of antiquark $'2'$. $l$ is the string length and $\omega$ is the 
orbital angular velocity of quark. $\hat{n}$ is the unit vector in the 
direction of the area sweep out by the rotational motion of string. Similarly 
we can calculate the orbital color magnetic moment of antiquark $'2'$.
The net color magnetic moment of a meson is obtained by summing up the 
color magnetic moments produced by quark as well as antiquark. Therefore the 
net color magnetic moment of a meson is given by 
\begin{eqnarray}
\label{meson} 
\vec{\mu}^a=&\frac{1}{2}&Q^al^2\omega\frac{m_2-m_1}{m_2+m_1}\hat{n}+[f^{abc}Q^b\{A^c_o(\vec{r}_1)-\vec{v}_1\cdot\vec{A}^c(\vec{r}_1)\}m^2_2\nonumber\\
&-&f^{ade}Q^d\{A^e_o(\vec{r}_2)-\vec{v}_2\cdot\vec{A}^e(\vec{r}_2)\}m^2_1]\frac{\pi l^2}{(m_2+m_1)^2}\hat{n}  
\end{eqnarray} 

According to eq\ref{meson} the color magnetic moment of a meson will have 
abelian as well as non-abelian components. Both abelian and non-abelian 
components are functions of masses of quarks. For a meson with same kind of 
quark/antiquark (like $c\overline{c}$, $b\overline{b}$ etc) the abelian 
component of color magnetic moment will be zero. 

\subsection{Color magnetic moment of baryons} 
A baryon consists of three quarks (say $q_1,q_2,q_3$). It can have three 
possible configurations described by fig 1. b,c,$\&$ d. On repeating the 
analysis of mesons we calculate the color magnetic moments of all these 
configurations. The color magnetic moment for configuration decribed in 
fig1.b is given by 
\begin{eqnarray}
\label{baryon-1} 
\vec{\mu}^{(1b)a}=&\frac{1}{2}&l^2\omega Q^a_1\frac{m_2+m_3-m_1}{m_1+m_2+m_3}\hat{n}+[f^{abc}Q^b_1\{A^c_o(\vec{r}_1)-\vec{v}_1\cdot\vec{A}^c(\vec{r}_1)\}(m_2+m_3)^2 \nonumber\\
&-&f^{ade}Q^d_1\{A^e_o(\vec{r}_2)-\vec{v}_2\cdot\vec{A}^e(\vec{r}_2)\}m^2_1]\frac{\pi l^2}{(m_1+m_2+m_3)^2}\hat{n} 
\end{eqnarray} 

Similarly color magnetic moments for other two possible configurations can be 
calculated. All these configurations are equally probable. Their average will 
be the color magnetic moment of a baryon. Therefore the color magnetic moment 
of a baryon is given by 
\begin{eqnarray}
\label{baryon} 
\vec{\mu}^a=&\frac{1}{6}&\frac{l^2\omega}{m_1+m_2+m_3}[Q^a_1(m_2+m_3-m_1)+Q^a_2(m_1+m_3-m_2)\nonumber\\
&+&Q^a_3(m_1+m_2-m_3)]\hat{n}+\left[[f^{abc}Q^b_1\{A^c_o(\vec{r}_1)-\vec{v}_1\cdot\vec{A}^c(\vec{r}_1)\}(m_2+m_3)^2\right. \nonumber\\
&-&f^{ade}Q^d_1\{A^e_o(\vec{r}_2)-\vec{v}_2\cdot\vec{A}^e(\vec{r}_2)\}m^2_1]\nonumber\\
&+&[f^{ab'c'}Q^{b'}_2\{A^{c'}_o(\vec{r}_1)-\vec{v}_1\cdot\vec{A}^{c'}(\vec{r}_1)\}(m_1+m_3)^2\nonumber\\
&-&f^{ad'e'}Q^{d'}_2\{A^{e'}_o(\vec{r}_2)-\vec{v}_2\cdot\vec{A}^{e'}(\vec{r}_2)\}m^2_2]\nonumber\\
&+&[f^{ab''c''}Q^{b''}_3\{A^{c''}_o(\vec{r}_1)-\vec{v}_1\cdot\vec{A}^{c''}(\vec{r}_1)\}(m_1+m_2)^2\nonumber\\
&&\left.-f^{ad''e''}Q^{d''}_3\{A^{e''}_o(\vec{r}_2)-\vec{v}_2\cdot\vec{A}^{e''}(\vec{r}_2)\}m^2_3]\right]\frac{\pi l^2}{(m_1+m_2+m_3)^2}\hat{n} 
\end{eqnarray} 

From the above expression it is clear that the color magnetic moment of 
baryons also have abelian as well as non-abelian components. 

\subsection{Chromomagnetic properties of nuclear matter} 
Let us consider a sample of nuclear matter where hadrons (irrespective of 
baryon or meson) do not interact among themselves. To study the thermodynamic 
properties we first calculate the partition function. The partition function 
for such system is given by
\begin{eqnarray}
z=exp\{-\beta(\epsilon-\vec{\mu}^a\cdot\vec{B}^a)\}+exp\{-\beta(\epsilon+\vec{\mu}^a\cdot\vec{B}^a)\}\ 
\end{eqnarray} 
where $\epsilon$ is the non-magnetic energy associated with a hadron, $\beta$ 
is the temperature inverse of hadronic system, and $\vec{B}^a$ is the external 
color magnetic field acting on the nuclear matter. Therefore the mean color 
magnetisation (color magnetic moment per unit volume), $\vec{M}^a$, of a 
hadronic medium is given by 
\begin{eqnarray}
\vec{M}^a=C\mu^atanh({\beta \mu^b B^b})
\end{eqnarray} 
where $C$ is number density of hadron for a stable nuclear matter but for 
unstable quark matter like quark-gluon plasma (QGP) there will be continuous 
production of hadrons then $C$ will indicate hadron production rate and 
obviously $\vec{M}^a$ will show the rate of change of magnetization of 
hadronic matter. 

At very large temperature 
\begin{eqnarray}
\label{charge}
\vec{M}^a=C\frac{\mu^a\mu^b B^b}{k_BT}
\end{eqnarray} 
On following the works of Karsch \cite{karsch} and Huang and Lissia 
\cite{huang} one can show 
\begin{eqnarray}
\label{hua} 
Q^2(T)=4\pi/\left[9 ln \left(\frac{T}{0.1254T_c}\right)^2\right]
\end{eqnarray} 
for three quark flavors. In eq\ref{hua} $T_c$ is the QCD phase transition 
temperature. In other words the color charge of a quark is a temperature 
dependent quantity. Therefore from equations \ref{meson},\ref{baryon},
\ref{charge},and \ref{hua} it is clear that the color magnetism of hadronic 
matter does not follow the Curie law. Due to the presence of non-linear 
temeprature dependence and non-abelian component of color magnetic moment it 
will show complicated behavior. 

\section{Conclusions}
So far we have studied chromomagnetism produced by quarks only. However 
hadrons are color neutral particles. But they may have non-zero chromomagnetic 
moment. Due to this chromomagnetic moment they will show color interactions 
among each other. Due to the non-abelian nature of strong interactions there 
are two components of chromomagnetic moment of hadrons: (i) abelian and (ii) 
non-abelian. This may cause the existence of different chromomagnetic phases 
of nuclear matter. From the analysis for a non-interacting hadronic system we 
have seen that chromomagnetism of hadronic matter does not follow the Curie 
law.  

\begin{acknowledgements}
The authors thank the anonymous referees for their constructive comments 
which helped to improve the presentation of this work. AR is thankful to 
IIT Ropar for their hospitality where some part of this work was done. The 
authors acknowledge the financial support from DST, India 
(grant no. SR/S2/HEP-13/2006). We are grateful to the people of India for 
their generous support for research. 
\end{acknowledgements}


\begin{thebibliography} {05}

\bibitem{georgi}
A. De R\`{u}jula, Howard Georgi, and S. L. Glashow,
Phys. Rev. D {\bf 12} 147(1975).

\bibitem{bourdeau}
Michele Bourdeau and Nimai C. Mukhopadhyay,
Phys. Rev. Lett. {\bf 58} 976(1987).

\bibitem{aramayo}
Oscar A. Rondon-Aramayo,
Nucl. Phys. A {\bf 490} 643(1988).

\bibitem{sapirstein}
Jonathan Sapirstein,
Phys. Rev. D {\bf 20} 3246(1979).

\bibitem{ninomiya}
M. Ninomiya and N. Sakai,
Nucl. Phys. B {\bf 190} 316(1981).

\bibitem{kapusta}
J. I. Kapusta,
Nucl. Phys. B {\bf 190} 425(1981). 

\bibitem{wong} 
S. K. Wong,
Nuovo Cimento A{\bf 65} (1970) 689.

\bibitem{karsch}
F. Karsch, 
Proceedings of the Workshop on Color confinement and Hadrons,
edited by H. Toki {\it et al}, (World Scientific, Singapore 1995) pp. 109. 

\bibitem{huang} 
S. Huang and Marcello Lissia, 
Nucl. Phys. B {\bf 438} 54(1995). 

\end{thebibliography}
\end{document}